\documentclass[conference]{IEEEtran}
\IEEEoverridecommandlockouts
\usepackage{cite}
\usepackage{amsmath,amssymb,amsfonts}
\usepackage{algorithmic}
\usepackage{graphicx}
\usepackage{textcomp}
\usepackage{xcolor}
\usepackage{amsmath}
\usepackage{amssymb}

\usepackage[title]{appendix}

\usepackage{multirow}
\usepackage{subcaption}
\usepackage{makecell}
\usepackage{float}
\usepackage{hyperref} 
\newcommand{\oursys}{CoPlay}

\begin{document}

\title{\oursys: Audio-agnostic Cognitive Scaling for Acoustic Sensing}



\author{\IEEEauthorblockN{Yin Li, Bo Liu, Rajalakshmi Nandakumar}
\IEEEauthorblockA{\textit{Cornell University}\\
\{yl3243, bl685, rajalakshmi.nandakumar\}@cornell.edu}
}

\maketitle

\begin{abstract}

Acoustic sensing manifests great potential in various applications like health monitoring, gesture interface, by utilizing built-in speakers and microphones on smart devices. However, in ongoing research and development, one problem is often overlooked: the same speaker, when used concurrently for sensing and other traditional audio tasks (like playing music), could cause interference in both, making it impractical to use. The strong ultrasonic sensing signals mixed with music would overload the speaker's mixer. To confront this issue of overloaded signals, current solutions are clipping or down-scaling, both of which affect the music playback quality, sensing range, and accuracy. To address this challenge, we propose \oursys, a deep learning-based optimization algorithm to cognitively adapt the sensing signal and run in real-time. It can 1) maximize the sensing signal magnitude within the available bandwidth left by the concurrent music to optimize sensing range and accuracy and 2) minimize any consequential frequency distortion that can affect music playback.
We design a custom model and test it on common types of sensing signals (sine wave or Frequency Modulated Continuous Wave FMCW) as inputs alongside various agnostic types of concurrent music and speech. First, we micro-benchmark the model performance to show the quality of the generated signals. Secondly, we conducted 2 field studies of downstream acoustic sensing tasks on 2 devices in the real world. A study with 12 users proved that respiration monitoring and gesture recognition using our adapted signal achieve similar accuracy as no-concurrent-music scenarios, whereas baseline methods of clipping or down-scaling manifest worse accuracy. A qualitative study also justifies that \oursys{} leaves music untouched, unlike clipping or down-scaling that degrade music quality.
\end{abstract}

   





\section{Introduction}

\begin{figure*}
    \begin{minipage}{0.645\textwidth}
        \centering
        \includegraphics[width=0.95\linewidth]{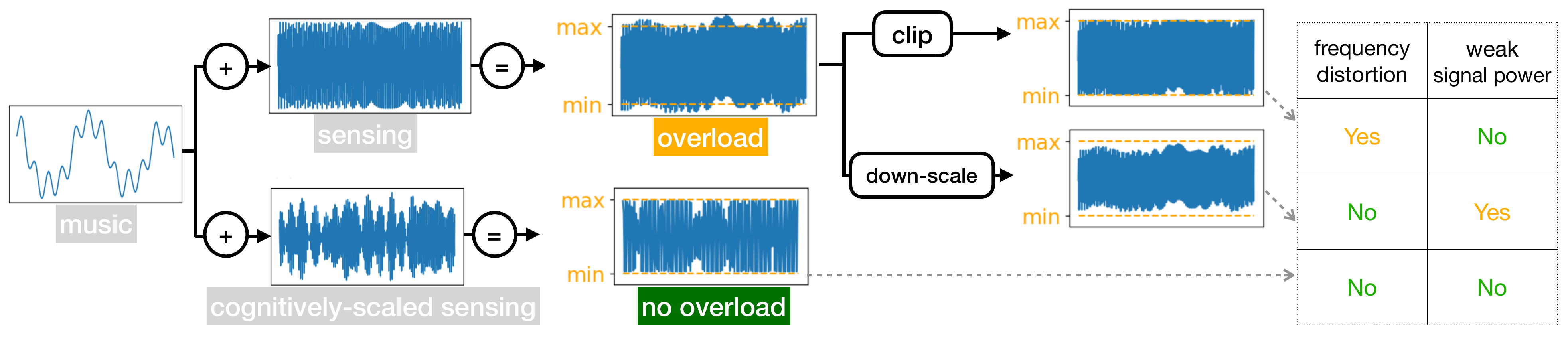}
        \vspace{-5pt}
        \caption{With concurrent music, our cognitive scaling avoids mixer overload, while clipping and down-scaling cause frequency distortion or weaken signal power.}
        \label{fig:teaser}
    \end{minipage}
    \hfill
    \begin{minipage}{0.335\textwidth}
       \centering
        \includegraphics[width=.98\linewidth]{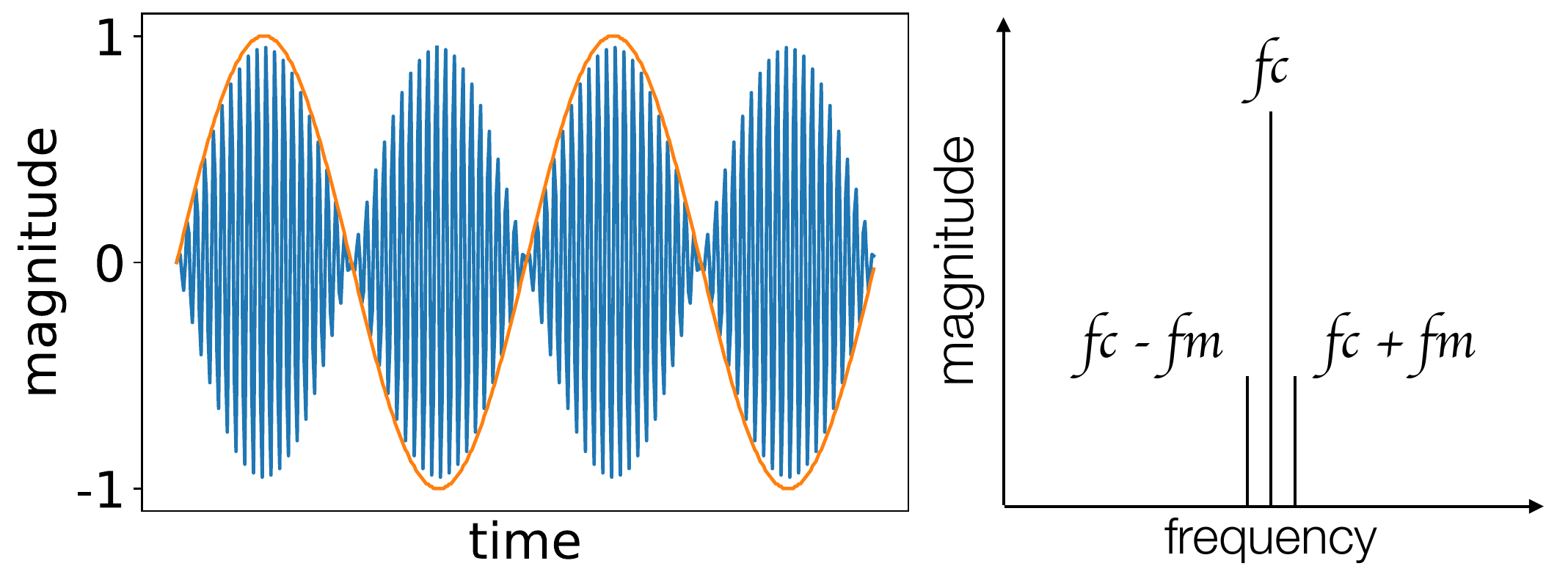}
        \vspace{-5pt}
        \caption{Inordinate scaling in time domain causes distortion in frequency domain, taking Amplitude Modulation as an example.}
    \label{fig:amdistortion}
    \end{minipage}
    \vspace{-15pt}
\end{figure*}

Acoustic sensing is an active research field, typically leveraging the built-in acoustic sensors on commercial-off-the-shelf (COTS) smart devices: it emits inaudible custom ultrasound signals from the speaker and then records the reflections using microphones. This transforms the device into an active SONAR system, capable of motion detection for various applications such as human gesture recognition and tracking~\cite{amaging2022, koteshwara2023presence, wang2016device, nandakumar2016fingerio}, biosignal monitoring~\cite{nandakumar2015contactless, wang2019contactless, song2020spirosonic}.
Acoustic sensing offers unique advantages as a novel perception modality: 1) Unlike cameras, it works even in non-line-of-sight or poor light conditions, providing enhanced privacy preservation. 2) The speed of sound is many orders of magnitude slower than the speed of radio frequency (RF), allowing for millimeter-level sensing resolution with a significantly narrower bandwidth. 3) They harness the increasing number of built-in speakers and microphones in IoT devices without necessitating hardware modifications for sensing. It has been actively studied per improving accuracy, security, etc. 

However, one critical issue in ongoing research and developments is unaddressed: most systems assume the speaker solely plays the sensing signal. Yet, speakers are conventionally used for audio applications such as music playback. Concurrently running music and sensing can overload the speaker's mixer. To elaborate, a speaker/audio mixer is a software component in the operating systems that mixes audio streams from different applications, via resampling, scaling, and adding them into a single output stream. If the mixed signal exceeds DAC's maximum numerical range, overload occurs.

In consequence, the mixer would take two possible actions: clipping off the overloaded part or down-scaling the whole to stay within range, as in Fig.~\ref{fig:teaser}. Currently, these two strategies are widely adopted by operating systems: Android, Windows, and Linux uses clipping; IOS and MacOS use scaling. ~\cite{li2022experience, koteshwara2023presence} identified that both methods will degrade the quality of music or sensing. 1) Clipping results in distortion in the frequency domain, leading to buzzing noise and affecting the performance of spectrum-based sensing algorithms such as FFT. 2) Down-scaling preserves frequency domain but significantly decreases transmitted signal power, lowering music play volume and weakening sensing detection range.

Beyond identifying the essence of the concurrent music problem in ~\cite{li2022experience, koteshwara2023presence}, some work tangentially address this using closed-form solutions like echo cancellation~\cite{koteshwara2023presence} or interleaving~\cite{nandakumar2017covertband}. But they are limited to certain tasks or hurts the signal quality and sensing performance. In this work, we aim to propose an algorithmic solution that outperforms prior closed-form methods.
So, we propose \oursys, a scaling algorithm, to dynamically adapt sensing signals in consideration of concurrent music applications and maintain optimal sensing and music playback. In specific, it can simultaneously 1) maximize magnitude within available bandwidth left by concurrent music, enabling optimal sensing performance 2) minimize any consequential frequency distortion.

Achieving these both goals simultaneously is hard because tweaking the signal amplitude in time domain introduces frequency domain distortion.
For example, amplitude modulation(AM) creates noisy side bands beside the carrier frequency. Our dynamic scaling of amplitude is worse.
Therefore, a deep learning algorithm is necessary to dynamically scale the signal, by formalizing these two objectives as an optimization problem. 
We name it \textit{cognitive} scaling model, inspired by cognitive radar~\cite{bell2015cognitive}, where the transmission signal(TX) is adaptively adjusted based on environmental changes via real-time perception-action cycles. In our paper, we dynamically adjust the Tx ultrasound according to the concurrent music, running in the speaker mixer to avoid overload while preserving both sensing and music quality.

Specifically, the cognitive scaling neural network should handle agnostic music and various types of sensing signals (regardless of the receiver-side sensing algorithms). So, in this work, we examine our model with two common sensing signals, sine wave and FMCW chirp, along with four types of concurrent audio like piano and podcast. 
Sine wave is simple, narrow-band, but prone to interference, while FMCW chirps~\cite{nandakumar2015contactless, zhou2018echoprint, li2022lasense} offers better correlation and flexible range resolution but are more complex in frequency spectrum thus more challenging for our model to generate accurately.
We had a workshop paper\cite{coplayworkshop2023}(\textit{anonymized}) presenting preliminary results of only sine waves to embark on prior research conversation. But it couldn't extend to FMCW signals. In contrast, \oursys{} incorporates advanced algorithms with broader experiments, allowing it to process both sine and FMCW.

To mitigate these challenges, we design a cognitive scaling neural network with custom layers, taking music and sensing as input and cognitively scaled sensing signal as output. Specifically, a custom link function with music as bias can filter and normalize inputs within dynamic bandwidth. A custom kernel further alleviates low-frequency noise stemming from the spectral bias and aliasing in convolution layers\cite{zhang2019making, rahaman2019spectral}. The loss terms also integrate the two optimization goals to maximize the amplitude and minimize frequency distortion.
To evaluate the model performance beyond numerical loss, we also conduct field studies in the real world to show how the adapted signal works in downstream tasks. The studies of respiration detection and gesture recognition with 12 users show that our adapted signals perform as well as in no-music scenarios. Besides, qualitative study manifests that we have better quality than clipped or down-scaled signals.
To summarize, our work has the following contributions.
\begin{itemize}
    \item We formalize the concurrent music issue as an optimization problem and design a cognitive scaling deep learning model for the mixer, ensuring optimal sensing performance and leaving concurrent agnostic audio untouched.
    \item Our extensive evaluation covers complex types of sensing signals including sine and FMCW and microbenchmarking under various ablation studies and datasets.
    \item The custom layers in our model offer valuable insights for the special scenario of non-auto-regressive modeling on ultra-high-frequency signals.
    \item Beyond model evaluation, we deploy the algorithm in real-world field studies with 12 users. The quantitative and qualitative results in downstream tasks of respiration detection and gesture manifest our great performance, while clipping and down-scaling hurt the performance.
\end{itemize}


\section{Related work}
As a pilot algorithmic solution to the concurrent music issue in acoustic sensing based on optimization, this section contains a rundown of this field and associated methodologies. First, we overview existing acoustic sensing research, noting some studies have offered preliminary investigations of the concurrent music issue, and some tangentially addressed it with closed-form fixes. Secondly, we elaborate on inspirational methods drew from other fields, including non-autoregressive approaches in audio generation, and anti-aliasing techniques in neural networks for computer vision.

\textbf{Acoustic sensing} is typically subjected to analysis measuring the Doppler effect\cite{gupta2012soundwave, koteshwara2023presence}, time of flight (ToF)~\cite{li2022lasense}, direction of arrival (DOA)~\cite{mao2019rnn, garg2021owlet}, etc.
The signal is usually custom-modulated, such as sine waves or FMCW. For example, Amazon Echo~\cite{koteshwara2023presence}, LLAP~\cite{wang2016device}, and ~\cite{mao2019rnn, fu2022svoice} uses high-frequency sine waves for presence detection, fingertips tracking, hand tracking and silence speech recognition. Cat~\cite{mao2016cat},  EchoPrint~\cite{zhou2018echoprint}, ApneaApp~\cite{nandakumar2015contactless} and Lasense ~\cite{li2022lasense} transmit FMCW signal for localization, face or respiration detection. Besides, synthetic aperture radar (SAR) algorithms are widely used for acoustic imaging, as adapted in ~\cite{mao2018aim, amaging2022}.
These research usually focus on improving the sensing accuracy without considering concurrent music. However, this problem could distort or weaken the received sensing signal, which hurts the phase-based/frequency-based sensing methods like DOA and SAR, specifically their convolution property (when using FFT or correlation). This underscores the reason behind our incorporation of FFT within the loss function.
As for the external background noise, which is more often mentioned in these works, we can easily filter it out with a high-pass filter. However, internal noise—like music playing from the same speaker can bring unique challenges, as highlighted above.

\textbf{The concurrent music problem:} Covertband~\cite{nandakumar2017covertband}, a human activity detection system using acoustics, first proposed to interleave the concurrent music with the sensing signal; but this method hurts the sample rate exponentially, which degrades both music and sensing, especially ultrasound (17kHz+) that requires at least double sample rate (34kHz) to avoid aliasing.
Both ~\cite{li2022experience, koteshwara2023presence} underscored that the concurrent music distortion results from mixer overflow and clipping.
The experience paper~\cite{li2022experience} validated the performance of their algorithm~\cite{li2022lasense} under down-scaling, without proposing new solutions to the problem. But as we discussed above, scaling down causes additional issues.
The Google patent ~\cite{koteshwara2023presence} tried to remove the concurrent music distortions in the spectrum by acoustic echo cancellation (AEC) component, assuming motions are non-symmetric in the spectrum but distortions are. However, it only applies to simple tasks of sensing such as Doppler-effect-based presence detection.
Therefore, our prior workshop paper\cite{coplayworkshop2023} formalized this problem as optimization, which is better than down-scaling or clipping. However, this preliminary paper was limited to sine waves, excluding FMCW signals. In contrast, \oursys{} introduces advanced techniques with broader experiments, enabling processing for both sine and FMCW. The increased complexity in frequency necessitates the utilization of \oursys's customized model.


\textbf{Waveform modeling:} Our objectives involve simultaneously maximizing the magnitude and minimizing consequential frequency distortion. 
This needs to be formalized as an optimization problem, making deep learning a window function. So, it is close to waveform modeling in the general audio domain.
Non-autoregressive (non-AR) raw audio generation is an active domain. Their sequence-to-sequence model architecture design for raw waveform has proved effective in many tasks such as including speech synthesis\cite{oord2016wavenet}, sound source separation\cite{stoller2018wave}, and music generation\cite{goel2022s}. Standard sequence modeling approaches like CNNs\cite{oord2016wavenet}, RNNs\cite{mehri2016samplernn}, and Transformers\cite{child2019generating} are adapted for waveform modeling. Particularly, WaveNet~\cite{oord2016wavenet} and its variants\cite{engel2017neural, peng2020non} are fundamental backbone components in such architectures. In this work, we adopt this backbone and show the limitations and comparison of other architectures through experiments.
Moreover, our task differs from common waveform modeling: the latter focuses on audible sound below a few hundred Hz; while our ultrasonic sensing signals are 17kHz+, posing unique challenges due to neural networks' spectral bias toward learning low-frequency features\cite{rahaman2019spectral}. Some studies attribute this bias to aliasing, caused by passing the activations through down-sampling layers like convolution and max pooling. Nevertheless, these investigations per spectral bias in neural networks are primarily in computer vision tasks\cite{zhang2019making, michaeli2023alias}, whose definition of high frequency is much lower than 17kHz. Inspired by these works, we design custom kernels to alleviate this issue in ultrasound modeling particularly.

\section{Method}
In this section, we first interpret \oursys's challenges of dynamic scaling and formalize our objectives as an optimization problem. Then, we introduce the model with custom layers.

\subsection{Understanding the effect of dynamically scaling the signal}
\label{sec:amplitudemodulation}

Achieving our two objectives simultaneously is hard because tweaking signal amplitude in time domain will distort frequency domain.
To understand the effect of inordinately tweaking the amplitude, consider amplitude modulation (AM) as an interpretable example, a telecommunications technique to transmit information by varying carrier wave amplitude. In Fig.~\ref{fig:amdistortion}, the high-frequency carrier is multiplied by the low-frequency base signal, i.e. amplitude modulated, making the frequency spectrum spread out with noisy side bands. 
In this case, the change is closed-form; side frequencies are $f_{c}+f_{m}$ and $f_{c}-f_{m}$ (Appendix \ref{sec:appendixAM}).
Building upon this observation, our cognitive scaling is namely multiplying the signal with a more complex window function, leading to various sidebands in the frequency. 
Therefore, this effect of altering signals necessitates our utilization of deep learning; closed-form signal processing or traditional machine learning can hardly dynamically control the distortion during generation.

\subsection{Sensing signal modulation types}

Acoustic sensing commonly uses two types of signals: sine waves and FMCW chirps, which serve different purposes in various applications.
\textbf{1) Sine waves} are single-frequency oscillation that varies with time according to the mathematical function $A \sin(2\pi ft + \phi)$, where $A$ is the amplitude of the sine wave, $f$ is the frequency, $\phi$ is the initial phase.
They are common for tasks such as signal testing, calibration, and reference signals, providing a well-defined waveform, easy to generate and analyze.
\textbf{2) FMCW chirps} have frequency changes linearly over time, basically repeated chirps defined as $A\sin(2\pi( f_{a}t + \frac{f_1-f_0}{2T}t^2))$, where $f_0$ and $f_1$ are the starting and ending frequency; $T$ is the duration per chirp.
They are widely used in radar and sonar, offering improved range resolution for precise distance measurements and multi-target discrimination.


\subsection{Define the problem as optimization}
\label{sec:problemdefinition}

To begin, we establish a formalization of the model input, output, and training objectives in loss function.

\textbf{Input and output:} Let $x$ be the sensing signal, i.e. a high-frequency sine or chirp. Concurrently, an App plays music $z$. Then the mixer mixes them within the maximum range of $[-m, m]$ allowed by the number of DAC bits. In this paper, we use signed 16-bit PCM, resulting in a range of $-2^{15}\sim2^{15}-1$. For computation convenience, we set $m = 2^{15}-1$.
In general, our goal is to generate a cognitively-scaled sensing signal $\hat{x}$ conditioned on the music $z$; and the $\hat{x}$ is optimal when satisfying the following objectives defined in the loss.

\textbf{Loss:} The first objective is to maximize the sensing amplitude within the bandwidth left by music. In other words, the amplitude of the mixed signal $\hat{x}+z$ in time domain should be close to the maximum allowed amplitude $m$. To restate, it is to minimize loss $q(\hat{x} + z, m)$, namely the difference between the magnitude of them.
However, this would cause consequential frequency distortion as explained in section~\ref{sec:amplitudemodulation}, which deteriorates the receiver-side algorithms since they generally rely on accurate frequency (either FFT convolution or correlation convolution). Therefore, we define another loss $p(x, \hat{x})$ as the difference between $x$ and $\hat{x}$ in frequency domain; it is the difference between their L-2-norm of FFT coefficients.
In summary, our goal is to minimize these two loss terms simultaneously as an optimization problem.

Next, we can further customize the loss for sine and chirps. The L-2 norm of FFT coefficients is then rearranged as 
\begin{align}
    p(x, \hat{x}) = 1 - \|\hat{c_{f_c}}\|_{2} + \frac{1}{N-1}\sum_{i \in -{f_c}}\|c_i-\hat{c_i}\|_{2}
\end{align}
where $c$ are the magnitude of the complex number coefficients in FFT, $FFT(x) = (c_i, ...)$, $FFT(\hat{x}) = (\hat{c_i}, ...)$. And $f_c$ is the target frequency bin containing the ultrasound frequency. $N$ is half of the window size since we only take the one symmetric half of the FFT output.
The magnitude is normalized by the window size and the maximum allowable amplitude of input $x$, i.e. $m$. The normalization guarantees that the two variable terms in $p$ are standardized and unbiased. 
We call the first half \textbf{target-loss} and the second half \textbf{recovery-loss}. The intention is to encourage a large value in target frequency bins and a close-to-zero value in the others.
Secondly, $q$ is the L-2 norm of $\hat{x}+z$ in time domain normalized by $m$, named as \textbf{amplitude-loss}.
\begin{align}
    q(\hat{x}+z, m) = 1 - \frac{1}{2N}\|\frac{\hat{x}+z}{m}\|_{2}
\end{align}
For FMCW chirp, the multiple frequency bins also need even amplitude to correlate well. So, an additional \textbf{variance-loss} for FMCW is built via standard deviation with zero correction:
\begin{align}
    s(\hat{x}) = \sqrt{\frac{\sum_{i \in {f_c}}(\hat{c_i} - \bar{\hat{c_i}})^2}{N_{f_c}-1}}
\end{align}
Finally, the overall loss equals to the weighted summation of these two terms, where $\alpha, \beta, \gamma$ are hyperparameters.
\begin{align}
    loss_{sine} &= \alpha * p(x, \hat{x}) + \beta * q(\hat{x}+z, m) \\
    loss_{chirp} &= \alpha * p(x, \hat{x}) + \beta * q(\hat{x}+z, m) + \gamma * s(\hat{x})
\end{align}
We will ablate study the hyperparameters in section~\ref{sec:benchmark}.
Note that, our loss may not guarantee that $\hat{x} + z$ falls within [$-m, m$]. Therefore, we need to embed this constraint into model layers, as introduced in the following section~\ref{section:model} per customized link function.
In practice, we split the sensing signal and music into small windows and run optimization for each window. For chirps, the window size should match the chirp size. In this paper, we use 18KHz sine wave and 18k-20kHz FMCW signal, with a window size of 512 and a sampling rate of 48KHz. These are common settings balancing efficiency and performance. And our model can be extended to other settings as long as training and testing settings match.

\subsection{Cognitive scaling neural network}
\label{section:model}

\begin{figure*}
    \begin{minipage}{0.52\textwidth}
        \centering
        \includegraphics[width=0.98\linewidth]{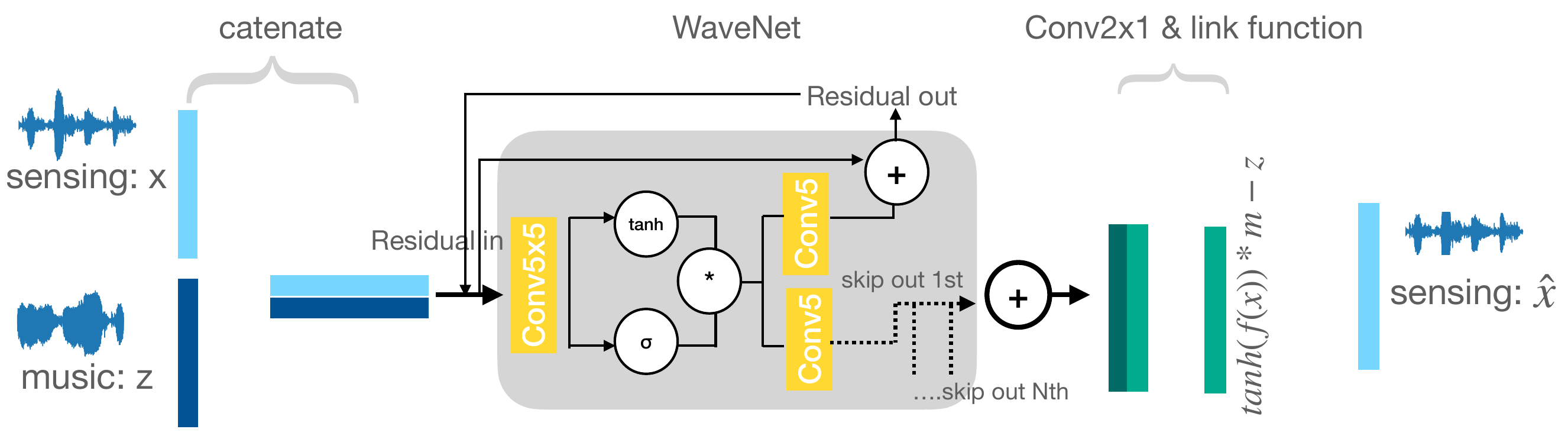}
        \vspace{-5pt}
        \caption{Cognitive scaling neural network.}
        \label{fig:modelArch}
    \end{minipage}
    \hfill
    \begin{minipage}{0.46\textwidth}
       \centering
        \begin{subfigure}[t]{.39\linewidth}
            \includegraphics[width=\linewidth]{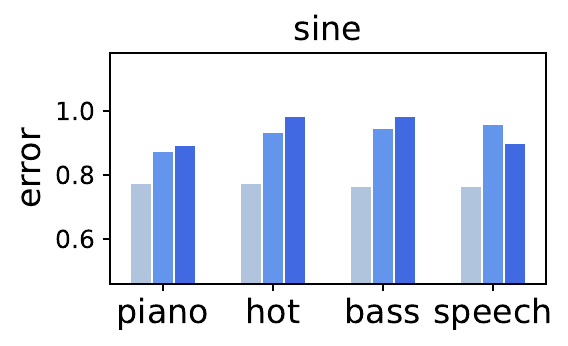}
        \end{subfigure}
        \begin{subfigure}[t]{.59\linewidth}
            \includegraphics[width=0.95\linewidth]{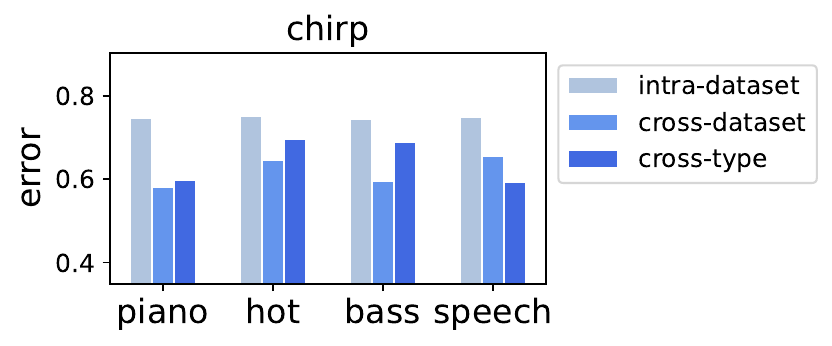}
        \end{subfigure}
        \vspace{-15pt}
        \caption{The model generalizes well to unseen datasets and new types of music or speech.}
        \label{fig:cross-dataset}        
    \end{minipage}
    \vspace{-15pt}
\end{figure*}

Next, we establish a model using a non-AR seq-to-seq network. As depicted in Fig.~\ref{fig:modelArch}, the inputs are music $z$ and sensing $x$ and the output is adapted sensing $\hat{x}$. The model architecture derives from a non-causal WaveNet backbone, followed by an anti-aliasing \textit(sinc) kernel and a custom link function for normalization with agnostic music as bias.

\textbf{Model architecture with WaveNet backbone.}
To elaborate, it combines a non-auto-regressive sequence-to-sequence structure with WaveNet backbones. While WaveNet architecture was originally developed for AR modeling, it has since become a basic backbone of various audio models. And we validate this choice in the micro-benchmark section by experimental comparison with sample CNN and RNN backbones. In detail, it employs a stack of dilated convolutional layers which provides better reception fields than conventional ones, making it adept at raw waveform modeling and processing sequential data more efficiently than RNNs. The input of the model is concatenated sensing signal $x$ and music $z$ of the same length of window size 512. Then the 2-row tensor is passed through 2 layers of Wavenet with fused addition and $tanh$ + multiplication, then dilated convolution (hidden size of 2, kernel size of 5, and dilation rate of 1), and finally residual skip layer (double hidden size except the last layer). 
Then the outputs pass through conv1d layer to merge the two rows, followed by custom layers to be described in the following sub-sections. Based on observation, we choose a large learning rate of 0.1 with \textit{Adam} optimizer that is stable in training.
Our implementation refers to DiffSinger~\cite{liu2022diffsinger} on HuggingFace.

\textbf{A sinc kernel for anti-aliasing.}
In early experiments, we observe non-negligible low-frequency noise in the model output, especially when modeling FMCW chirps. It is a common problem in computer vision community that the neural network learns low frequency faster and also down-scaling in layers like convolution and max pooling could induce aliasing. 
Therefore, we design a custom \textit{sinc} kernel for band-passing middle activations to isolate specific frequency bands of interest within the signal.
The unique properties of the \textit(sinc) filter, notably its brick-wall characteristic in the frequency domain 
as its Fourier transform is a rectangular function (Appendix ~\ref{sec:appendixSinc}),
makes it a preferred choice of filters over Hamming, etc.
So, by putting \textit(sinc) kernel as a layer before the final link function, it mitigates artifacts that may arise in the first stage.
For implementation, we use \textit{sinc\_impulse\_response} function in the Pytorch experimental nightly version. The negative of it is a high-pass filter. And they can sum to one band-pass filter or more.
To demonstrate its effectiveness, we perform ablation in the micro-benchmark section. 
Another advantage is its potential to serve as a parameterized kernel when building models for new types of sensing signals other than chirp or sine. A parameterized cutoff could learn a dynamic filter for even multiple signals trained in one model. While we currently train standalone models dedicated to each sensing signal, so that each model achieves optimal performance since we focus on substantiating and validating the underlying concept.

\textbf{A link function for normalization with agnostic-music as bias.}
As described, $\hat{x} + z$ should be within the range of [$-m, m$], namely a normalization with agnostic music as bias. To restate, the output $\hat{x}$ should be $a \times m + z$ where $a$ is the activation from the previous layer within [0, 1]. 
So, to embed this constraint as a trainable layer, we leverage the concept of link function in Generalized Linear Models (GLM); It is in analogy to the logit function for the logistic regression, which uses the prior to target on the probability of binary classification ranging from 0 to 1. Similarly, \oursys{} uses a hyperbolic tangent function ($tanh$) as link function that scales the output within [-1, 1].
A bonus is that it encourages the value to approach -1 or 1, which is useful in our cases because we want to maximize the magnitude of the waveform, i.e. close to 1 or -1. 
In contrast, clipping simply truncates the output, resulting in loss of information or under-fitting, while $tanh$ is more effective especially when data is continuous and has a wide range of values.
So, we design a $tanh$ layer rather than a min-max clipping. $\hat{x} = tanh(a) * m - z = \frac{e^a - e^{-a}}{e^a + e^{-a}} * m - z$ where $a$ represents the activations from the previous layer. This scales $\hat{x} + z$ within $[-m, m]$. 
For inference, we feed $\hat{x} + z$ into the mixer in case $\hat{x}$ exceeds the numerical range.
Finally, to quantify the output as integers while the model is trained in float64, we employ a workaround using the \textit{round} function: it detaches the difference and then adds it back. Otherwise, \textit{round} or \textit{typecast} are not differentiable in PyTorch.

\section{Micro-benchmark}
\label{sec:benchmark}
This section details the experiment setup and various model performance micro-benchmarks, including different datasets, music types and volume. Ablation studies also justify our design in model architecture and multi-task learning weights. 
Code is published on Github\footnote{https://github.com/lydhr/CoPlay}.

\textbf{Evaluation metric} in this section is mainly the loss, as defined in section~\ref{sec:problemdefinition}.
It indicates the relative performance among micro-benchmarks. Furthermore, to offer interpretable analysis of output quality, we provide visualizations for frequency spectrograms and cross-correlation ranging profiles.
In the next section, we will further assess the performance in downstream field studies like gesture and respiration detection.

\subsection{Datasets}
We use four audio types, each with two datasets, to evaluate generalization across agnostic audio. They are cut into equal-length clips, then paired with same-length sensing clips as defined above.
\textbf{1) Piano:} One dataset is the Beethoven dataset~\cite{mehri2016samplernn}, consisting of 10-hour piano sonatas, often serving as a benchmark music dataset for audio generation. We interpolate the data to a sample rate of 48kHz and cast the 8-bit quantization into a 16-bit signed integer. 
Another dataset is the YouTubeMix dataset~\cite{youtubemix} with 4-hour higher-quality recordings, which we format in the same way as the Beethoven dataset. The following are also downloaded via YoutubeMix toolset adhering to the same format.
\textbf{2) Hot songs:} We select two non-stop playlists of hot songs, featuring top-ranking Billboard songs from 2019 and 2022\cite{ds-pop2019, ds-pop2022}, spanning various genres like Pop, Rock, and Hip-hop. Both are around two hours long.
\textbf{3) Bass: } To ensure the inclusion of low-frequency audio, we had two bass-centric playlists~\cite{ds-bassedm, ds-basslofi}, lasting for one hour and two hours.
\textbf{4) Speech:} Besides musical audio, we add two conversational speech datasets: a podcast featuring Selena Gomez\cite{ds-podcastSelena} and a Conan O'Brien podcast\cite{ds-podcastConan}. Both are over one hour.

\subsection{Experiment results}

\textbf{Generalization across datasets and music types:} In total, we have 8 datasets. We define \textit{intra-dataset} experiment as training and testing the model with the same dataset, \textit{cross-dataset} as training and testing with same-type different datasets, and \textit{cross-type} as with different types of datasets (the most out-of-distribution case). 
Note that, we avoid randomly splitting the dataset for testing that is non-realistic in usage. Instead, the last 10\% sequentially is testing set. The rest is then randomly split into 80\% training and 20\% validation. All trainings run 10 epochs to converge unless otherwise specified.
Fig.~\ref{fig:cross-dataset} shows the average error for sine and chirps separately, grouped by music types.
For better plots, the y-axis is log-scaled testing loss multiplied by $1e5$.
In summary, the results demonstrate strong generalization, as the errors in cross-dataset and cross-type are similar. Intra-dataset is the best as expected because the distribution is close between testing and training. However, for chirp data, intra-dataset is counter-intuitive, and it takes 10x more epochs to converge. This probably owes to the complexity of chirp that needs more training data to learn effectively, while intra-dataset uses only \~60\% of the data available in cross-dataset/type setup.
Notably, speech also achieves robust cross-type results, highlighting great generalization ability, as the most distinguishable type.
Overall, errors remain low, which are further justified by downstream tasks in following section.


\begin{table}[H]
    \centering
    \begin{tabular}{lllll}
     \hline
     & \makecell{target\\-loss} & \makecell{recovery\\-loss} & \makecell{amplitude\\-loss} & \makecell{variance\\-loss} \\
     \hline
    sine & 0.5928 & 0.0012 & 0.9558 & - \\
    chirp & 0.1852 & 0.0013 &  0.9560 & 0.0214 \\
    \hline
    \end{tabular}
    \caption{The value of sub-loss terms.}
    \vspace{-15pt}        
    \label{table:individualLoss}
\end{table}

\textbf{Decompose the loss:} The overall error comprises multiple loss terms as defined in section~\ref{sec:problemdefinition}. Hence, we decompose it into individual loss term values to assess each corresponding objective. Table.~\ref{table:individualLoss} is the average sub-loss on hot songs datasets. They are training loss including three sub-loss values for sine and an additional variance-loss for FMCW as defined above. By default weights, each sub-loss ranges from 0 to 1. 
We observe that target-loss and amplitude-loss are the bottlenecks, aligning with the intuition that minimizing frequency distortion and maximizing the amplitude are opposed objectives, which is a common issue in multi-task learning. This table serves as a reference for tweaking the weights of loss terms to specific downstream task requirements. For instance, near-field sensing tasks can sacrifice amplitude for minimal distortion.

        

\begin{figure}
    \centering
    \begin{subfigure}[t]{.47\linewidth}
        \centering
        \includegraphics[width=0.82\linewidth]{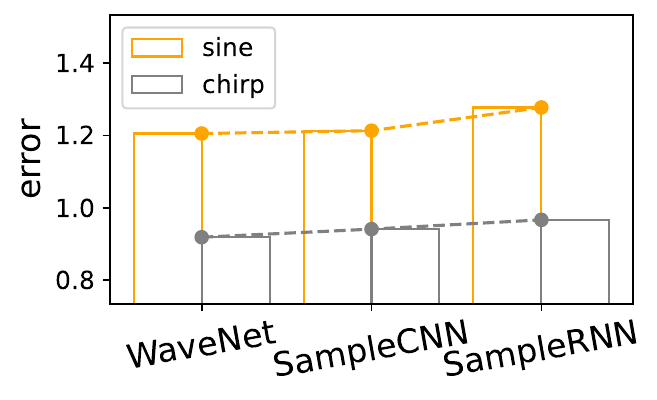}
        \vspace{-5pt}
        \caption{model architecture}
        \label{fig:abl_modelArch}
    \end{subfigure}
    \begin{subfigure}[t]{.51\linewidth}
        \centering
        \includegraphics[width=0.71\linewidth]{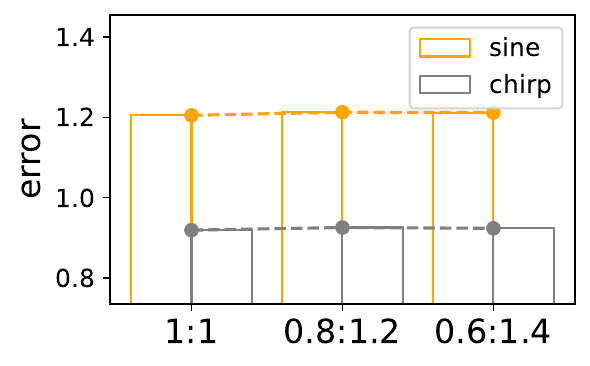}
        \vspace{-5pt}
        \caption{sensing volume: music volume}
        \label{fig:abl_musicvolume}
    \end{subfigure}
    \begin{subfigure}[t]{.49\linewidth}
        \centering
        \includegraphics[width=0.8\linewidth]{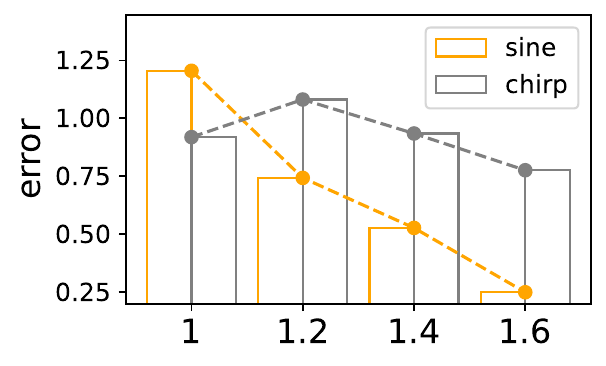}
        \vspace{-5pt}
        \caption{weight of recovery-loss}
        \label{fig:abl_hyperparam}
    \end{subfigure}
    \begin{subfigure}[t]{.49\linewidth}
        \centering
        \includegraphics[width=0.79\linewidth]{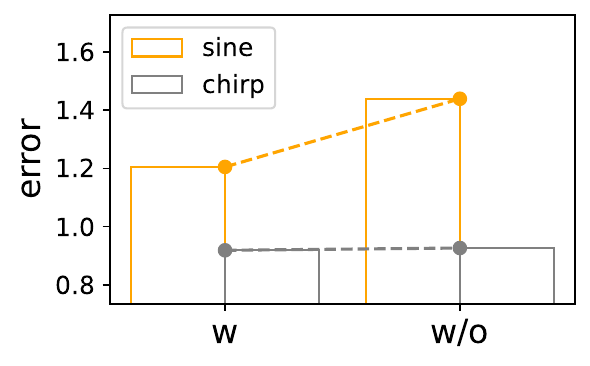}
        \vspace{-5pt}
        \caption{with or without sinc kernel}
        \label{fig:abl_filter}
    \end{subfigure}
    \vspace{-5pt}
    \caption{Ablation study results.}
    \vspace{-5pt}
\end{figure}

\textbf{Varying music volume: } Although a high music volume might degrade our model performance, Fig.~\ref{fig:abl_musicvolume} shows that the overall error does not increase with music volume.
In detail, we train with the same sub-loss experiment and test with increasing levels of music volume. The x-axis indicates the ratio of sensing signal volume and music volume. The default ratio is 1:1, and 0.6:1.4 is when music is roughly double in power, equivalently 7.3db louder.
So, \oursys{} is stable when training with agnostic music even in unseen volume, likely due to the normalization in amplitude-loss and recovery-loss. 

\textbf{Ablation of model architecture:} To justify our choice, we compare three models backbones namely CNN, RNN, and Wavenet with the same number of recursive layers. Fig.~\ref{fig:abl_modelArch} shows that Wavenet slightly outperforms CNN and both surpass RNN. But CNN has a limited reception field so we chose Wavenet,  which improves this with dilated convolution layers. Furthermore, the sample RNN uses LSTM layers, making it less efficient to train.

\textbf{Ablation of multi-task loss terms:} Our loss function comprises sub-losses with equal default weights, which can be adjusted to emphasize specific objectives (as in ~\ref{sec:problemdefinition}). So, we test increasing recovery-loss weight from 1 to 1.2, 1.4, and 1.6 to prioritize minimizing frequency distortion, while the other weights are 1 by default; this is a practical use case when certain sensing systems, like near-field tasks, can compromise the power for minimal frequency distortion. Fig.~\ref{fig:abl_hyperparam} shows error decreasing as weight increases, providing a reference for tuning this ratio for the downstream tasks in practice.

\textbf{Ablation of sinc filter: } To validate the efficacy of the \textit{sinc} filter, Fig.~\ref{fig:abl_filter} exhibits improved performance with the presence of \textit{sinc} kernel, in both the sine and chirp models, especially the former. We attribute this enhancement to the narrow target main lobe in the spectrum of sine, thus more side lobes are to be suppressed. The \textit{sinc} kernel helps significantly reduce these lobes, resulting in a lower numerical loss. So, we incorporate this essential layer in all experiments.

\subsection{Spectral analysis}

\begin{figure}
    \centering
    \begin{subfigure}[t]{.37\linewidth}
        \includegraphics[width=\linewidth]{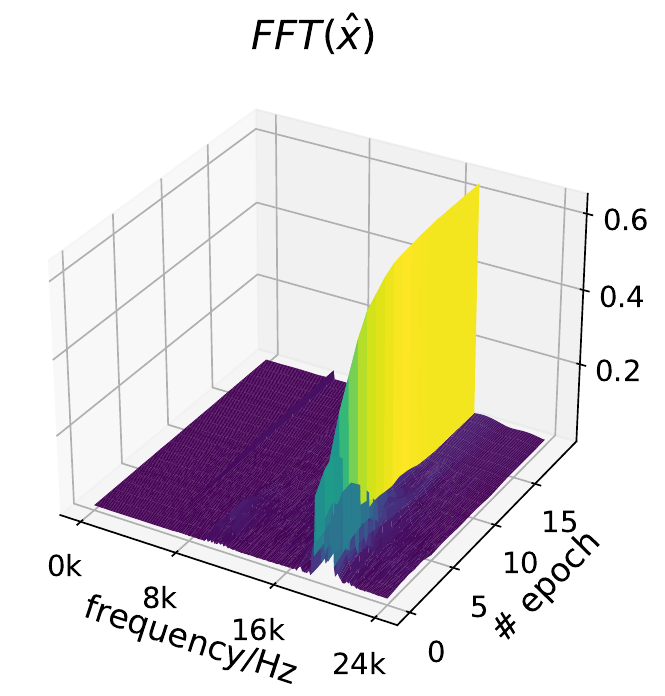}
        \caption{sine}
    \end{subfigure}
    \begin{subfigure}[t]{.37\linewidth}
        \includegraphics[width=\linewidth]{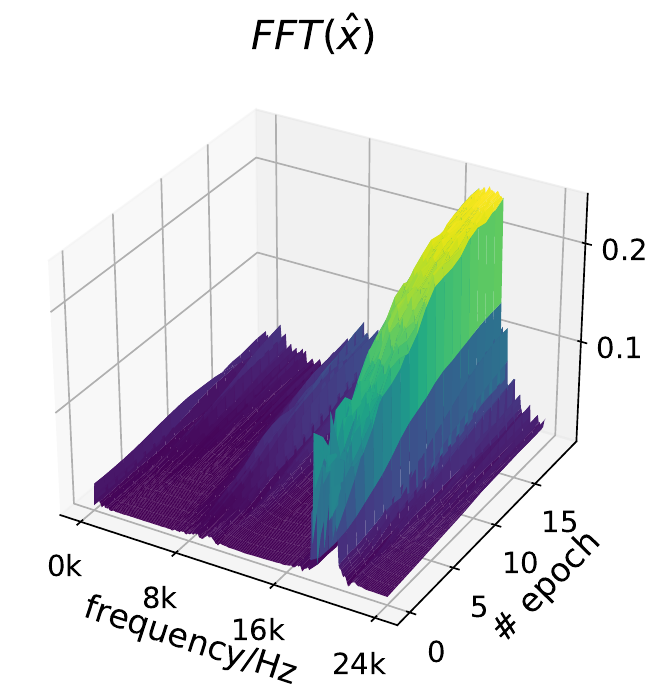}
        \caption{chirp}
    \end{subfigure}
    \vspace{-5pt}
    \caption{Generated signal $\hat{x}$ grows as expected as training epochs increase: target frequency rises, noises are suppressed.}
    \label{fig:3Dloss}
    \vspace{-15pt}
\end{figure}

\textbf{Training procedure in 3D frequency domain: } Although the numerical loss can indicate the model performance, we seek a more comprehensive perspective by visualizing the model output in frequency spectrum. This helps confirm that the output aligns well with expectations.
In Fig.~\ref{fig:3Dloss}, we take the FFT of a sample window output $\hat{x}$ and screenshot it with the increasing training epochs. The rise of the yellow peaks indicates that it converges to an 18kHz sine wave or 18k-20kHz FMCW as the training epoch increases. The side lobes are much weaker than the main lobes, which manifest good SNR. (Moreover, we further substantiate the SNR by qualitative study in section~\ref{sec:fieldstudy})

\begin{figure*}
    \begin{minipage}{0.32\linewidth}
        \centering
        \includegraphics[width=0.98\linewidth]{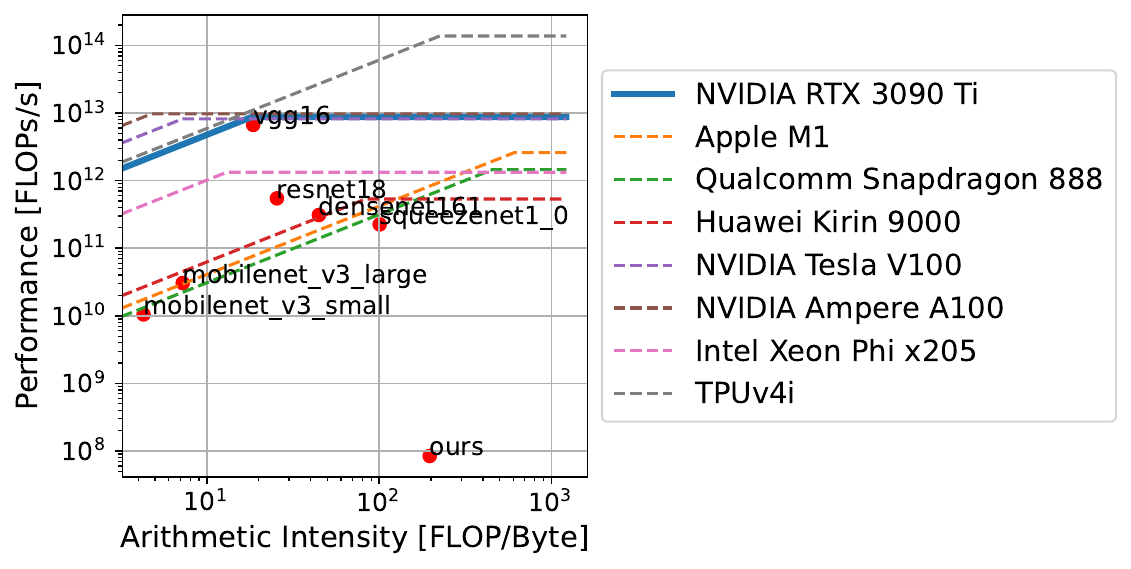}
        \vspace{-5pt}
        \caption{Roofline plot.}
        \label{fig:roofline}
    \end{minipage}
    \hfill
    \begin{minipage}{0.66\linewidth}
        \centering
        \includegraphics[width=0.98\linewidth]{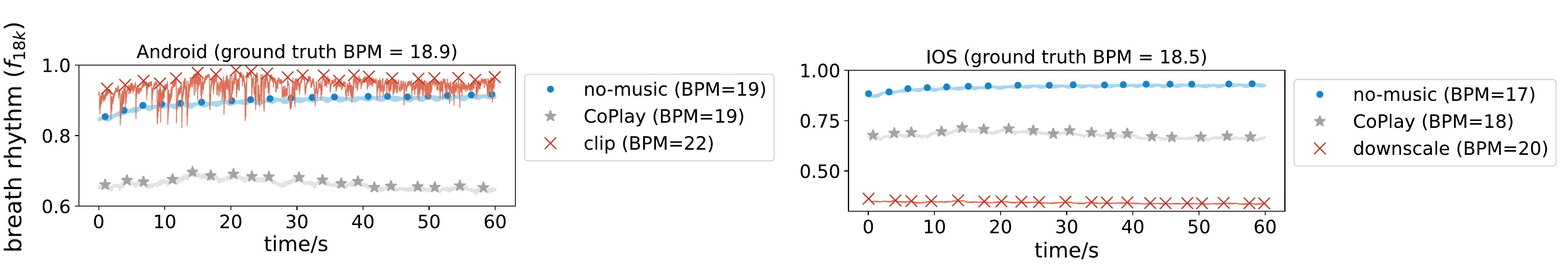}
        \caption{The amplitude of detected breath rhythm of the three scenarios: no-music, \oursys, and clip(Android) or downscale(IOS).}
        \label{fig:br_vis_device}
    \end{minipage}
    \vspace{-18pt}
\end{figure*}

\begin{figure}[H]
    \centering
    \vspace{-10pt}
    \includegraphics[width=0.8\linewidth]{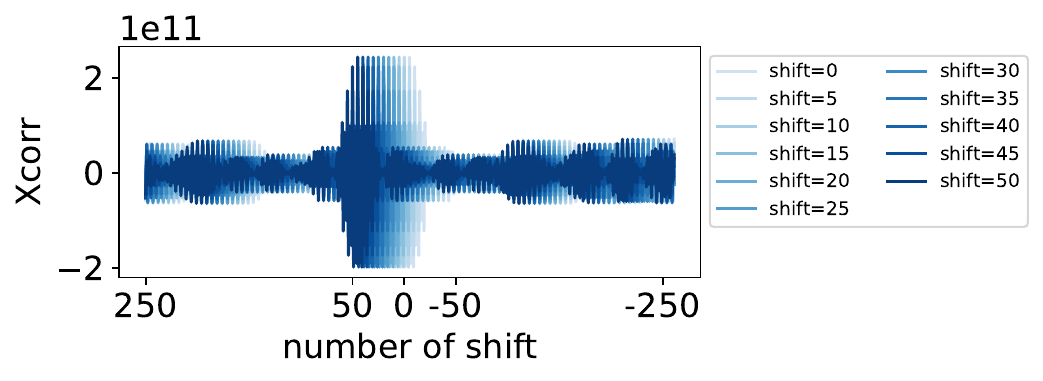}
    \vspace{-5pt}
    \caption{Generated signal has good cross-correlation property: the distinct peak moves with change of shifts.}
    \vspace{-10pt}
    \label{fig:xcorr}
\end{figure}

\textbf{Ranging profile with cross-correlation:}
Likewise, certain sensing algorithms employ cross-correlation (Xcorr) to obtain the ranging profile instead of utilizing FFT. Nevertheless, they basically use equivalent computation components: convolution or even DFT. For instance, an accelerated computation of Xcorr is to perform a frequency-domain Xcorr, which is faster than that in the time domain~\cite{xCorrFD}, because shifting a signal in the time domain is equivalently scaling it in the frequency domain.
Therefore, in addition to plotting the FFT spectrums of the generated signal, Fig.~\ref{fig:xcorr} shows the Xcorr correlation output for the cognitively-scaled signal. By rolling the signal in the time domain, the distinct peak in Xcorr output changes with the number of shifts as expected. In summary, this justifies that our adapted signal works well even if the receiver-side sensing algorithm uses Xcorr.

\subsection{Computation efficiency}
\oursys{} should function as a mixer algorithm for buffered audio stream in the speaker, processing small windows of audio clips. In this paper, window size is 512, common for acoustic sensing and efficiency.
Our model size is as compact as 5.5 KB, with forward/backward pass size of 0.98 MB, and an input size of 0.07 MB. The estimated total size of training is 1.05 MB, easily fitting into the RAM of commercial smartphones or other mobile devices during inference. Despite training on an RTX 8000 GPU, inference runs for roughly 625 fps on mobile devices, equivalently $1/625=1.6ms$ per frame. So, a buffer interval around $512/48000Hz = 10ms$ could leave quite enough time for inference on one frame, making it real-time with no perceptible delay.
Furthermore, the entire audio is pre-known by the speaker in many cases, so it can process the signal upfront before playing. 
Fig.~\ref{fig:roofline} also shows our superior efficiency compared to other popular models on RTX 3080; we even outperform mobile models like MobileNet, SqueezeNet, and DenseNet, due to our small dilated layers. It's equivalently 10X more power-efficient than background Apps. The dashed rooflines from other popular mobile SoCs(System-on-a-Chip) and GPUs compare the peak performance and memory bandwidth, validating our viability on most mobile chips, including Snapdragon 888 and Kirin 9000.
Moreover, literature on model compression and acceleration could further improve the efficiency, which is beyond the scope of this work.

\section{field study}
\label{sec:fieldstudy}

To validate \oursys's efficacy in real-world beyond the micro-benchmark, we run two downstream acoustic sensing tasks: respiration rate detection and hand gesture recognition, with concurrent audio. The study is across 2 tasks, 2 sensing signals, 3 audio types, and 2 device types with 12 users, compared against 3 baselines.
Experiments are controlled to isolate \oursys's effects, avoiding confounding factors like occlusion or range.
Moreover, since \oursys{} process signals before playback in speaker, downstream factors (like environmental interference, audible leakage and nonlinearity) wouldn't impact \oursys{} upstream and can be addressed via other existing methods such as filtering~\cite{li2022experience, wang2019contactless}.


\subsection{Field study tasks and evaluation metrics}
\textbf{1) Respiration monitoring} is a well-explored case, manifesting the ability of millimeter-level detection of acoustic sensing. It measures breath per minute (BPM) by detecting waveform peaks of chest movement during inhalations. So, our results report mean absolute error (MAE) of BPM alongside waveform visualizations.
\textbf{2) Hand gesture recognition} highlights motion recognition capabilities of acoustic sensing. Here we classify three gestures: swipe, push, and pinch, which are essential in gesture control systems. Our results report classification accuracy alongside the confusion matrix.

\subsection{Field study experiment setup}

\textbf{Hardware:} We test with two mobile phones: a Samsung S10 phone and an iPhone 13, representing Android and IOS, whose current solutions to mixer overflow are different: Android turns to clip, IOS turns to scale down. Phones are placed on the desk around 40cm away from the seated user facing front; one at a time.
For respiration detection, ground truth is from the Go Direct Belt with a force sensor~\cite{godirectPythonSDK}. It has a built-in breath rate calculation every 10 seconds with a window size of 30 seconds.
For gesture recognition, users follow a guided video displaying gestures every 3 seconds.

\textbf{Software:} We developed a custom data collection app for both Android and iOS. The recorder operates at 48kHz with 16-bit PCM, using a mono channel for both transmission and reception. Concurrent music plays at 50-60dB, aligning with typical mobile audio settings.
A laptop runs a data collection program for synchronized ground truth recording from the belt SDK~\cite{godirectPythonSDK} or gesture guiding video playback. It operates as a Flask RESTful API on a MacBook, triggered by the mobile app for synchronous sensing and ground truth logging.

\textbf{Sensing algorithm:} We use two classic receiver-side algorithms for sine and FMCW separately.
While various factors affect accuracy, including the sensing algorithm itself, our focus is on demonstrating \oursys's effectiveness relative to baselines, not optimizing the receiver-side algorithm.
1) For respiration rate detection, we use an 18kHz sine wave, extracting its frequency bin from the received signal, whose amplitude corresponds to chest-to-device distance; then we apply Scipy peak detection function with default parameters to estimate BPM. Besides, a band-pass filter (8-22Hz) before peak detection can filter out non-breathing motion like body rotation, since typical human breath rate is 12-18Hz. The analysis follows Vernier belt’s 30s window and 10s interval, with padding in initial 30s data.
2) For gesture recognition, we transmit FMCW of 18k-20kHz and then dechirp the received signal with cross-correlation to get a range profile. Each window subtracts its previous one to eliminate the static environment and let the motions stand out. To minimize interference from unintended nearby movements, users are instructed to stay still, isolating \oursys{}'s effectiveness since robustness under noise is not the focus of this study.
Finally, the range profile, as a single-channel 2D feature map, is input into a simple CNN classifier~\cite{cifarCNN}, which is a 2-conv-layer network with max pooling, ReLU, plus three linear layers. Then by training with Adam optimizer and cross-entropy loss, we got a 3-class classifier.
Besides, to avoid overfitting on our small dataset, we augment the feature map by adding Gaussian noise. 
Training follows a leave-one-user-out scheme, and reported results are the average accuracy.

\subsection{Study procedure}

Participants are first briefed on the IRB and study procedure without disclosing the audio order or content to play. Each user had 2 sessions (one for respiration, one for gestures), with 6 sub-sessions per session, totaling 12 minutes, i.e. over 2 hours for all 12 users.
Each session included three sub-sessions on Android phone: 1) sensing signal only, no-music, 2) audio + cognitively-scaled sensing, and 3) audio + sensing (causing clipping). Another three sub-sessions ran on iPhone similarly except that the third case will cause down-scaling instead. Sub-sessions 1 and 3 served as baselines to compare with \oursys{} in sub-session 2.
Audio clips were from datasets used in the micro-benchmark, featuring three 20-second samples of piano, hot songs, and speech.
Users started recording, switched phones and belt between sub-sessions, and then filled out qualitative questionnaires after every three sub-sessions.

\subsection{Field study results with baselines}

\begin{figure}
    \centering
    \begin{subfigure}[t]{.47\linewidth}
        \includegraphics[width=0.9\linewidth]{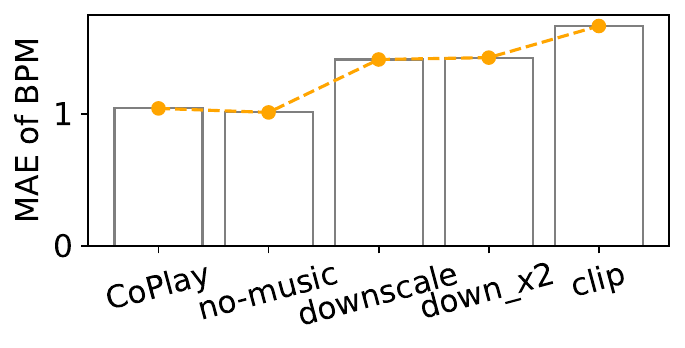}
        \caption{breath rate detection}
    \end{subfigure}
    \begin{subfigure}[t]{.4\linewidth}
        \includegraphics[width=\linewidth]{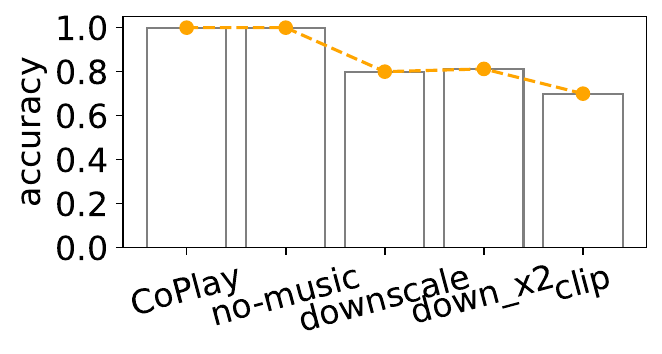}
        \caption{gesture recognition}
    \end{subfigure}
    \vspace{-5pt}
    \caption{Compared with sensing with no music, the baselines degrade sensing performance, while \oursys{} does not.}
    \vspace{-15pt}
    \label{fig:error_baseline}
\end{figure}

\textbf{\oursys{} V.S. Baselines (clipping and down-scaling).}
The four baselines are no-music baseline, clipping baseline, and two downscale baselines, measuring the MAE of BPM and gesture recognition accuracy. As depicted in Fig.~\ref{fig:error_baseline}, \oursys{} with concurrent music achieves similar results as the no-music baseline; while all the other three baselines exhibit inferior performance than the no-music. These findings are an average of across devices, users, and audio types.
In down-scaling, sensing power is reduced by half with the remaining half allocated to music. The down\_x2 further scales to 1/4 of the original, lowering volume by 12db (as sound pressure level is $20\log_{10}{power}$), then the music takes 3/4; it is also equivalent to double the sensing distance, as signal intensity is inversely proportional to the square of the distance from the source.
Overall, both clipping and downscaling underperform compared to \oursys{}. Clipping is the worst, even more than aggressive downscaling, while downscaling may degrade quadratically with increasing distance.
The next section will elucidate these observations by a visual breakdown.


\begin{figure}
    \centering
    \includegraphics[width=.65\linewidth]{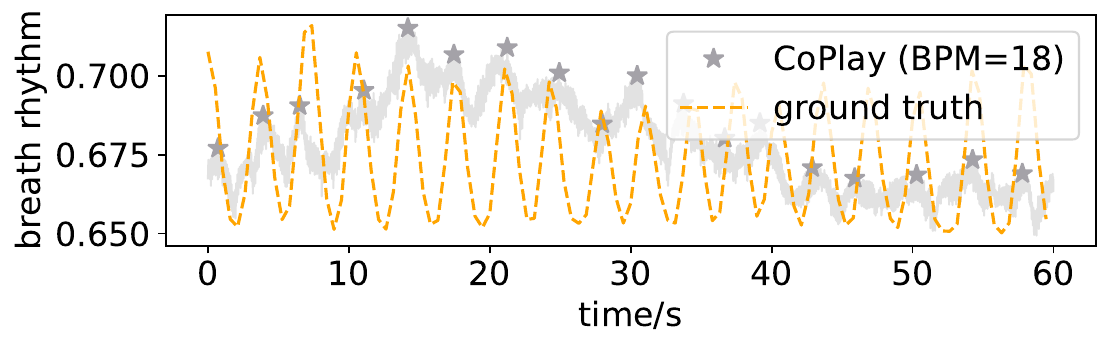}
    \vspace{-7pt}
    \caption{The detected breath of \oursys{} V.S. ground truth.}
    \label{fig:br_vis_groundtruth}
    \vspace{-15pt}
\end{figure}

\textbf{Visualization and error analysis.}
Fig.~\ref{fig:br_vis_device} depicts the amplitude of detected breath rhythm of the three scenarios: no-music, \oursys, and clip(Android) or downscale (IOS). To recap, the amplitude of breath rhythm is the normalized FFT amplitude at 18kHz, named $f_{18k}$, roughly proportional to the chest-to-device distance.
First, it is obvious that the no-music scenario for both Android and IOS is the strongest. While clipping maintains a high amplitude, it is significantly noisy due to distortion. In contrast, \oursys{} and no-music produce clear periodic waveforms.
Secondly, downscaling weakens the signal by nearly 50\%, while \oursys{} reduces it by only 15\%; Its crests and troughs are also less distinct, making peak detection ineffective.  
The ground truth BPM of the two selected samples are 18.9 and 18.5. As shown in figure legend, \oursys{} and the no-music baseline are close to the ground truth BPM while the others have a larger error. The dots along each line are the individual breath selected by the peak detection algorithm. 
Fig.~\ref{fig:br_vis_groundtruth} further zooms in on the first \oursys{} plot, overlaying it with the ground-truth dashed line (the normalized raw data from the force sensor). The peaks align well despite amplitude variations, validating \oursys's effectiveness.

\begin{figure}
    \centering
        \begin{subfigure}[t]{.48\linewidth}
            \includegraphics[width=0.8\linewidth]{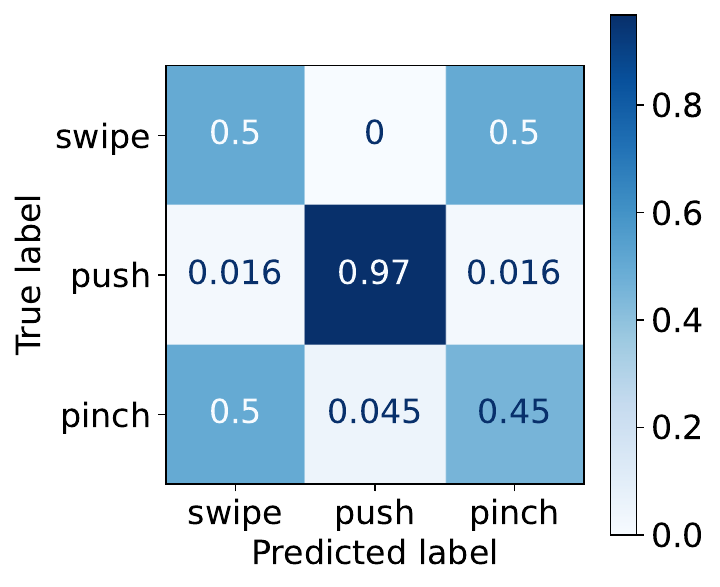}
            \vspace{-5pt}
            \caption{downscale}
        \end{subfigure}
        \hfill
        \begin{subfigure}[t]{.48\linewidth}
            \includegraphics[width=0.8\linewidth]{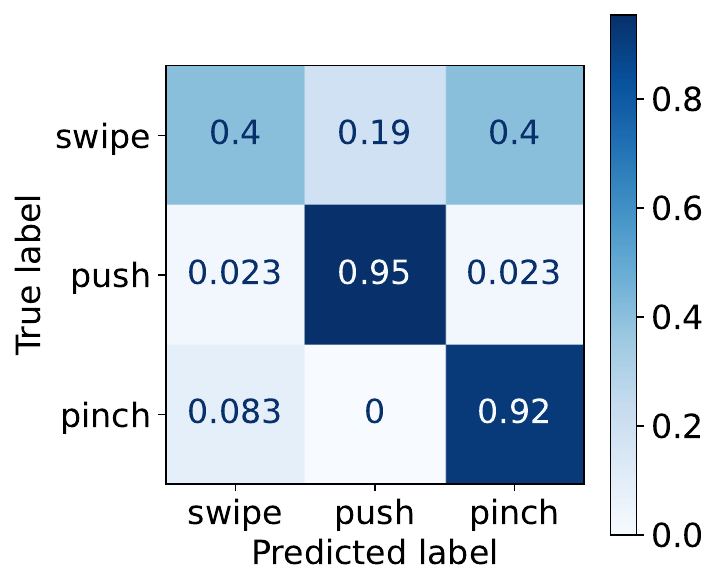}
            \vspace{-5pt}
            \caption{clip}
        \end{subfigure}
        \vspace{-5pt}
        \caption{Confusion matrix of gesture classification.}
        \vspace{-15pt}
        \label{fig:gesture_confusionMatrix}
\end{figure}

Similarly, we take a closer look at the errors of gesture classification using confusion matrix. Since \oursys{} and the no-music baseline achieve 100\% accuracy, Fig.~\ref{fig:gesture_confusionMatrix} only presents results of downscaling and clipping. As expected, wireless sensing is more sensitive to vertical motions like pushing, where reflection distance changes significantly. Thus, errors are higher for \textit{swipe} and \textit{pinch}, while \textit{push} is almost always accurately recognized.

\textbf{Qualitative study.}
Table.~\ref{tab:qualitative} is the qualitative study result, evaluating the playback quality of each sub-session per level of buzzing noise, loudness, and delays. The scores are the average of all users, in a 1-5 scale (integer) per question: \textit{Q1) Did you hear a buzzing noise? Q2) Did you hear the music at adequate volume? Q3) Did you perceive delay or discontinuity?}
In general, the results justify that \oursys{} leaves the music untouched. While Q1 manifests that clipping causes audible noise, basically buzzing.
Q2 shows that downscaling hurts audio power, making music less audible to users. To elaborate, in the study, we calibrated the system volume of the phone to 50db, a common loudness of music playback or a nearby conversational talk. The volume was measured by another smartphone App placed next to the study device. When downscaling happens, the music volume reduces by 6db per each half downscaling.
Q3 is to prove that the computation overhead of \oursys{} does not cause any discontinuity or delay between the chunks of processed clips.

\begin{table}[H]
\centering
\begin{tabular}{lccc}
\hline
 & clipping & downscale & \oursys \\ \hline
Q.1 ($\downarrow$5 = buzzing) & 4.72 & 1.13 & 1.22 \\
Q.2 ($\uparrow$5 = loud) & 4.14 & 1.12 & 4.49 \\
Q.3 ($\downarrow$5 = delay) & - & - & 1 \\
\hline
\end{tabular}
\vspace{-3pt}
\caption{Qualitative study results (1-5 scale): \oursys{} do not degrade music 
but clipping and downscaling do.}
\vspace{-10pt}
\label{tab:qualitative}
\end{table}
\section{discussion and future work}



This work explores 2 sensing signals in 2 applications. Future studies should explore a more expansive array of cases to thoroughly assess the versatility of the proposed approach. It should engage more intricate downstream tasks, like fine-grained gesture recognition, on diverse hardware.

We acknowledge a potential increase in computational costs, particularly for multiple channels. Although we assume multiple channels carry duplicate signals, 3D spatial audio may leverage slight variations requiring custom scaling and add to the computation load. Future work should evaluate this cost in realistic scenarios and explore acceleration methods like model compression and acceleration, also in comparison with other sensors' resource cost, like an MCU.


We aim to further explore speaker mixer behavior through real-world experiments on diverse devices, since factors like sensor count and placement might warrant investigation. Smart speakers like Amazon Echo Dot and Google Home, with their custom hardware and OS, may offer distinct insights into the concurrent music problem.




\bibliographystyle{IEEEtran}
\bibliography{sample-base}

\begin{appendices}
\section{An example of inordinate amplitude modulation and consequential frequency distortion.}
\label{sec:appendixAM}

Amplitude modulation (AM) is a modulation technique where the amplitude of a carrier signal is varied in proportion to the instantaneous amplitude of a modulating signal. Here is a simple case and the formulas of how the side frequencies derivative\cite{ampmodulationWiki}.

The carrier signal is denoted as \[  c(t) = A_c \cos(2\pi f_c t) \]

And the modulating signal is \[ m(t) = A_m \cos(2\pi f_m t) \]

The amplitude modulation (AM) signal results when carrier is multiplied by the positive quantity of the modulating signal:
\begin{align}
   y(t) = &[1 + \frac{m(t)}{A_c}] c(t) \\
   = &A_c \cos(2\pi f_c t)  \\
   &+ A_m \cos(2\pi f_c t) \cos(2\pi f_m t ) 
\end{align}

By rearranging
\begin{align}
&A_m \cos(2\pi f_c t) \cos(2\pi f_m t )\\
= &\frac{A_m}{2}[\cos(2\pi(f_c + f_m)t) + \cos(2\pi(f_c - f_m)t)] \\
= &\frac{A_c}{2}\frac{A_m}{A_c}[\cos(2\pi(f_c + f_m)t) + \cos(2\pi(f_c - f_m)t)] 
\end{align}

Finally, 
\begin{align}
y(t) = &  A_c \cos(2\pi f_c t) \\
&+ \frac{A_c m_a}{2}\cos(2\pi(f_c + f_m)t) \\
&+ \frac{A_c m_a}{2}\cos(2\pi(f_c - f_m)t) 
\end{align}
where $m_z = modulation index = \frac{A_m}{A_c}$.
By deriving $y(t)$, it becomes a modulated signal with two additional frequencies $f_c + f_m$ and $f_c - f_m$ besides the original $f_c$. 
The three components are called the carrier, the upper sideband (USB), and those below the carrier frequency constitute the lower sideband (LSB).
This is an explicit example of how inordinate changes in time domain amplitude result in side lobes in frequency distortion. Which guides the problem definition in this paper.

\section{Sinc function as a brick-wall filter.}
\label{sec:appendixSinc}

If we need a filter that yields minimal artifacts, its frequency response needs to be a rectangular function, namely a brick-wall filter:

\[ H(f) = \text{rect}(\frac{f}{2B}) =
\begin{cases}
0 & \text{if } |f| > B \\
\frac{1}{2} & \text{if } |f| = B \\
1 & \text{if } |f| < B
\end{cases}
\]
Where $B$ is the bandwidth, i.e. the cutoff frequency.
Then its impulse response is given by the inverse Fourier transform of its frequency response:
\begin{align}
    h(t) &= F^{-1}\{H(f)\} = \int_{B}^{-B} exp(2\pi i f t) df
\end{align}
Let $u = 2\pi i f t$, then
\begin{align}
    h(t) &= \frac{1}{2\pi i t} \int_{-2\pi i B t}^{2\pi i B t} e^{u} du\\
    &= \frac{1}{2\pi i t} (e^{2\pi i B t} - e^{-2\pi i B t})\\
    &= \frac{1}{2\pi i t} 2 i \sin(2\pi i B t))\\
    &= 2 B \frac{\sin(2\pi i B t)}{2\pi i B t})\\
    &= 2 B sinc(2Bt)
\end{align}

It turns out to be a $sinc$ function with cutoff $B$. 
Furthermore, \textit{windowed-sinc} finite impulse response is an approximation of \textit{sinc} filter. It is obtained by first evaluating \textit{sinc} function for given cutoff frequencies, then truncating the filter skirt, and applying a window, such as the Hamming window, to reduce the artifacts introduced from the truncation.


\section{METHODOLOGICAL TRANSPARENCY \& REPRODUCIBILITY APPENDIX}

As part of the Methodological Transparency \& Reproducibility Appendix (META), we will release our code for experiments on Github upon the accepting of this paper and the datasets are available online as described in the paper.

\end{appendices}

\end{document}